\documentclass[amsmath, amsfonts, superscriptaddress, showpacs, twocolumn, prb]{revtex4}
\usepackage{graphicx}
\usepackage{epsfig}
\usepackage{bm}
\usepackage{dcolumn}
\usepackage{amsmath}
\usepackage{amssymb}
\usepackage{euscript}
\usepackage{verbatim}

\begin{document}

\title{Interaction effects on thermal transport in quantum wires}

\author{Alex Levchenko}
\affiliation{Materials Science Division, Argonne National
Laboratory, Argonne, Illinois 60439, USA}
\affiliation{Department of
Physics and Astronomy, Michigan State University, East Lansing,
Michigan 48824, USA}

\author{Tobias Micklitz}
\affiliation{Dahlem Center for Complex Quantum Systems and Institut
f\"{u}r Theoretische Physik, Freie Universit\"{a}t Berlin, 14195
Berlin, Germany}

\author{Zoran Ristivojevic}
\affiliation{Laboratoire de Physique Th\'{e}orique-CNRS, Ecole
Normale Sup\'{e}rieure, 24 rue Lhomond, 75005 Paris, France}

\author{K.~A.~Matveev}
\affiliation{Materials Science Division, Argonne National
Laboratory, Argonne, Illinois 60439, USA}

\begin{abstract}
  We develop a theory of thermal transport of weakly interacting electrons in
  quantum wires. Unlike higher-dimensional systems, a
  one-dimensional electron gas requires three-particle collisions for energy relaxation. The fastest
  relaxation is provided by the intrabranch scattering of comoving electrons
  which establishes a partially equilibrated form of the distribution
  function. The thermal conductance is governed by the slower interbranch
  processes which enable energy exchange between counterpropagating
  particles. We derive an analytic expression for the thermal conductance of
  interacting electrons valid for arbitrary relation between the wire length
  and electron thermalization length. We find that in sufficiently long wires
  the interaction-induced correction to the thermal conductance saturates to an
  interaction-independent value.
\end{abstract}

\date{September 16, 2011}


\maketitle

\section{Introduction}

The classical Drude theory of electronic transport provides
universal relation between electric and thermal transport
coefficients known as the Wiedemann-Franz law,~\cite{Abrikosov}
\begin{equation}\label{WF}
K=\frac{\pi^2T}{3e^2}G\,.
\end{equation}
Here $K$ and $G$ are, respectively, the thermal and electric
conductances, $T$ is the temperature in energy units $(k_B=1)$, and
$e$ is the electron charge. The relation represented by
Eq.~\eqref{WF} is a natural one since for noninteracting particles
both charge and energy are carried by the electronic excitations.
Furthermore, as long as elastic collisions govern the transport, the
validity of the Wiedemann-Franz law has been confirmed in the case
of arbitrary impurity scattering.~\cite{Chester-Thellung} However,
an account of the electron-electron interaction effects within the
Fermi-liquid theory gives corrections to both $G$ and
$K$.~\cite{EE,Catelani} These lead to a deviation from the
Wiedemann-Franz law [Eq.~\eqref{WF}], which is associated with the
inelastic forward scattering of electrons. In general, violation of
the Wiedemann-Franz law is a hallmark of electron interaction
effects and thus is of conceptual interest.

In one-dimensional conductors, such as quantum wires or quantum Hall
edge states, the electron system can no longer be described as a
Fermi liquid but instead is expected to form a Luttinger
liquid.~\cite{LL} It has been shown that for the perfect Luttinger
liquid conductor, such as an impurity-free single-channel quantum
wire, interactions inside the wire neither affect conductance
quantization~\cite{MS-theorem}
\begin{equation}\label{G-0}
G_0=\frac{2e^2}{h}\,,
\end{equation}
nor change the thermal conductance of the system~\cite{FHK}
\begin{equation}\label{K-0}
K_0=\frac{2\pi^2T}{3h}\,.
\end{equation}
Since both $G$ and $K$ remain the same as in the case of
noninteracting electrons, the Wiedemann-Franz law \eqref{WF} holds
for an ideal Luttinger liquid conductor.

There are two important exceptions known in the literature. The
first one is a Luttinger liquid with an impurity studied by Kane and
Fisher.~\cite{Kane-Fisher} In that case electron backscattering
takes place which strongly renormalizes both $G$ and $K$, such that
the Wiedemann-Franz law is violated. The second case is the
Luttinger liquid with long-range inhomogeneities studied by Fazio
\textit{et al}.~\cite{FHK} If the spatial variations related to
these inhomogeneities occur on a length scale much larger than the
Fermi wavelength, electrons will not suffer any backscattering. The
electric conductance will, therefore, be given by its noninteracting
value [Eq.~\eqref{G-0}]. At the same time the thermal conductance
$K$ will be altered by interactions. The reason for this is as
follows. For the system with broken translational invariance,
momentum is not conserved. As a result, there are allowed certain
pair collisions which conserve the number of right and left movers
independently but provide energy exchange between them. These are
precisely the scattering processes that thermalize electrons and
thus lead to the violation of the Wiedemann-Franz law.

Recent advances in the fabrication of tunable constrictions in
high-mobility two-dimensional electron gases have allowed precise
and sensitive thermal measurements in clean one-dimensional systems.
These include experiments on the thermal transport of single-channel
quantum wires, where a lower value of the thermal conductance than
that predicted by the Wiedemann-Franz law was observed at the
plateau of the electrical conductance.~\cite{Schwab,Chiatti} Another
set of experiments reported enhanced thermopower in low-density
quantum wires~\cite{Nicholls,Sfigakis} and quantum Hall
edges.~\cite{Granger} Remarkable experiments based on
momentum-resolved tunneling spectroscopy provided direct evidence
for the electronic thermalization in one-dimensional
systems.~\cite{Birge,Altimiras,Barak} Clearly interaction effects
are responsible for the observed features; however, Luttinger liquid
theory does not provide an adequate description for these
observations.

Ongoing theoretical efforts in the study of one-dimensional electron
systems focus on nonequilibrium dynamics~\cite{GGM,Takei} and
consequences of the nonlinear dispersion in
transport~\cite{Lunde,JTK,TJK,ATJK,MAP,KOG,AZT} that lie beyond the
scope of conventional Luttinger liquid paradigm. The kinetics of
one-dimensional electrons with nonlinear dispersion is peculiar.
Indeed, constraints imposed by the momentum and energy conservation
allow either zero-momentum transfer or exchange of momenta for pair
collisions. Neither process changes the electronic distribution
function, and as a result they have no effect on transport
coefficients and relaxation. Therefore, three-particle collisions
play a central role.~\cite{Sirenko}

In this paper we study the fate of the Wiedemann-Franz law and the
origin of the electron energy relaxation in clean single-channel
quantum wires, accounting for the scattering processes that involve
three-particle collisions. The paper is organized as follows. In the
next section, Sec.~\ref{Sec-Preview}, we place our work into the
context of recent studies on equilibration in quantum wires and
explain the concept of partially equilibrated electron liquids,
which is central for our study. We then develop in Sec.~\ref{Sec-BE}
the theory of the thermal transport in one-dimensional electron
liquids based on the Boltzmann equation with three-particle
collisions included via the corresponding collision integral and
scattering rate. We elucidate the scattering processes involved,
discuss the role of the spin and limitations of our theory, and
summarize our findings in Sec.~\ref{Sec-Summary}. Additional
comments and directions for future work are presented in
Sec.~\ref{Sec-Discussion}. Various technical aspects of our
calculations are relegated to
Appendixes~\eqref{Sec-Appendix-PE}-\eqref{Sec-Appendix-l-Q}.

\section{Partially equilibrated one-dimensional electrons}\label{Sec-Preview}
\subsection{Noninteracting electrons}

\begin{figure}
  \includegraphics[width=8cm]{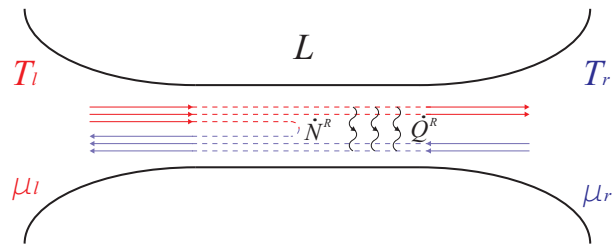}\\
  \caption{(Color online) Schematic picture of a quantum wire of length $L$.
    Electrons in the left and right leads are described by Fermi
    distribution functions characterized by temperatures $T_{l(r)}$
    and chemical potentials $\mu_{l(r)}$. Due to three-particle
    collisions electrons may backscatter and also exchange energy between
    the subsystems of warmer right movers and colder left movers.}\label{Fig1}
\end{figure}

Noninteracting electrons propagate ballistically through the wire.
They keep a memory of the lead they originated from and remain in
equilibrium with the corresponding lead electrons. If voltage ($V$)
and/or temperature differences ($\Delta T$) are applied across the
wire, right- and left-moving particles are at different equilibria,
and the distribution function takes the form
\begin{equation}\label{f-0}
 f_p =
\frac{\theta(p)}{e^{\frac{\varepsilon_p - \mu_{l}}{T_l}}+1}
          +\frac{\theta(-p)}{e^{\frac{\varepsilon_p -
          \mu_{r}}{T_r}}+1}\,,
\end{equation}
where $\varepsilon_p=p^2/2m$ is the energy of an electron with
momentum $p$ and $\theta(p)$ is the unit step function.
$T_l=T+\Delta T/2$, $T_r=T-\Delta T/2$, and  $\mu_l=\mu + eV/2$,
$\mu_r=\mu-eV/2$ are the different temperatures and chemical
potentials of left and right leads, respectively (see
Fig.~\ref{Fig1}). Employing the distribution function from
Eq.~\eqref{f-0} to linear order in $V$ and $\Delta T$, one readily
finds the electric and heat currents $I=G_0V|_{\Delta T=0}$ and
$I_Q=K_0\Delta T|_{I=0}$, with conductances $G_0$ and $K_0$, which
coincide with the earlier stated noninteracting values,
Eqs.~\eqref{G-0} and \eqref{K-0}.~\cite{Note}

In the presence of weak interactions the distribution \eqref{f-0}
describes an out-of-equilibrium situation. Collisions lead to
electrons exchanging energy and momentum, with some particles
experiencing backscattering. As a result, net particle ($\dot{N}^R$)
and heat ($\dot{Q}^R$) currents flow between the subsystems of
right- and left-moving electrons, relaxing $V$ and $\Delta T$ (see
Fig.~\ref{Fig1} for a schematic illustration). The effect of
electron-electron collisions on the distribution function depends
strongly on the length of the wire. Short wires are traversed by the
electrons relatively fast, leaving interactions only little time to
change the distribution in Eq.~\eqref{f-0} considerably. In the
limit of a very long wire, on the other hand, one should expect full
equilibration of left- and right-moving electrons into a single
distribution, even in the case of weak interactions. It turns out
that there exists a hierarchy of three-particle scattering
processes, classified by the corresponding relaxation rates or,
equivalently, inelastic scattering lengths, which have different
effects on the electron distribution function Eq.~\eqref{f-0}.

\subsection{Partially equilibrated electrons}

\begin{figure}
  \includegraphics[width=8cm]{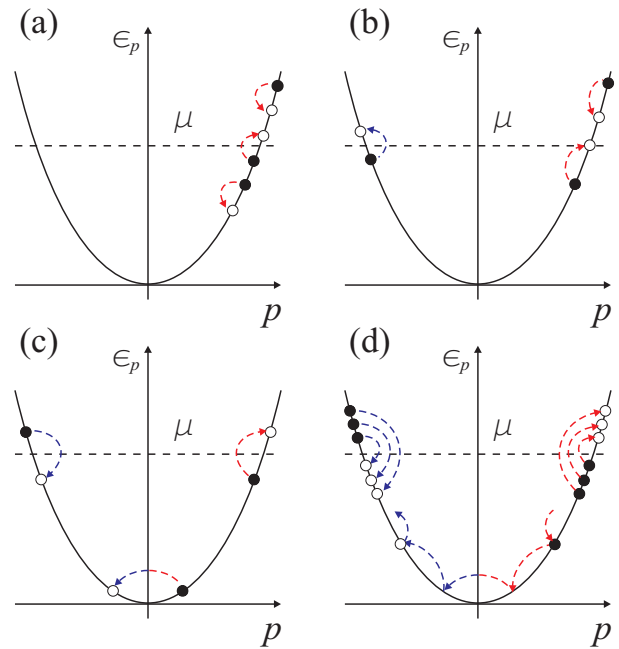}
  \caption{(Color online) (a) Intra-branch relaxation of co-moving electrons that establishes
  the partially equilibrated form of the distribution function
  in Eq.~\eqref{f-part-eq}. (b) Dominant interbranch three-particle process
  for the energy exchange $\dot{Q}^R$
  between counterpropagating electrons that contributes to the thermal
  conductance correction. (c) Leading three-particle collision which results
  in a finite rate $\dot{N}^R\propto e^{-\mu/T}$ and thus a temperature-dependent
  correction to the conductance of a short
  wire (Ref.~\onlinecite{Lunde}).
  (d) Equilibration mechanism: multistep diffusion through the bottom of the band
  of an electron from the right to the left Fermi point accompanied
  by the excitation of many electron-hole pairs
  (Refs.~\onlinecite{JTK,TJK,ATJK}).
  }\label{Fig2}
\end{figure}

We start our discussion of the different scattering processes
involved in the electronic relaxation with the process shown in
Fig.~\ref{Fig2}(a). This three-particle collision provides
\textit{intra}-branch relaxation within the subsystem of
right-moving electrons. A similar relaxation process for the left
movers is not shown in the figure but is implicit. This process can
be described by the corresponding inelastic scattering length, which
we denote in the following as $\ell_a$. The precise form of $\ell_a$
and its temperature dependence are model specific. They are
determined by the scattering amplitude for the given interaction
potential and the phase space available for this scattering
[Fig.~\ref{Fig2}(a)] to occur. Quite generally one may argue that
$\ell_a$ scales as a power of $T$. Indeed, at low temperatures,
$T\ll\mu$, all participating scattering states are located within
the energy strip $\sim T$ near the Fermi level. What is important
for the present discussion is that for wires with length
$L\gg\ell_{a}$ intrabranch electron collisions [Fig.~\ref{Fig2}(a)]
become so efficient that the initial distribution function
Eq.~\eqref{f-0} will be modified by interactions. One can find the
resulting distribution by employing the following observation.
Intrabranch collisions conserve \textit{independently} six
quantities. These are the number of right and left movers,
$N^{R/L}=\sum_{p\gtrless0}f_p$, their momenta,
$P^{R/L}=\sum_{p\gtrless0}pf_p$, and energies,
$E^{R/L}=\sum_{p\gtrless0}\varepsilon_pf_p$. The form of the
resulting electron distribution function $f_p$ can be obtained from
a general statistical mechanics argument by maximizing the entropy
of electrons, $S=-\sum_p[f_p\ln f_p+(1-f_p)\ln(1-f_p)]$, under the
constraint of the conserved quantities~\cite{TJK}
\begin{equation}\label{f-part-eq}
f_p
=\frac{\theta(p)}{e^{\frac{\varepsilon_p-pu^R-\mu^R}{\mathcal{T}^R}}+1}+
\frac{\theta(-p)}{e^{\frac{\varepsilon_p-pu^L-\mu^L}{\mathcal{T}^L}}+1}\,.
\end{equation}
This distribution is characterized by six unknown parameters
(Lagrange multipliers) which have a transparent physical
interpretation. Indeed, in Eq.~\eqref{f-part-eq} $\mathcal{T}^{R/L}$
are effective electron temperatures for right and left movers
different from those in the leads. The parameters $u^{R/L}$ have
dimension of velocity and account for the conservation of momentum
in electron collisions. Finally, $\mu^{L/R}$ are unequilibrated
chemical potentials of left- and right-moving particles. In
principle, all these parameters may depend on the position along the
wire.

For longer wires \textit{interbranch} three-particle collisions [see
Fig.~\ref{Fig2}(b)] become progressively more important. Unlike
intrabranch relaxation these processes allow energy and momentum
exchange between the subsystems of right and left movers; thus
$P^{R/L}$ and $E^{R/L}$ are no longer independently conserved.
However, the full momentum $P=P^R+P^L$ and energy $E=E^R+E^L$ are
obviously conserved. Corresponding to Fig.~\ref{Fig2}(b), the
scattering length $\ell_b$ is model specific and calculated in
Appendix \ref{Sec-Appendix-l-Q}. We note here that for all
interaction potentials we studied $\ell_{b}/\ell_{a}\sim \mu/T\gg1$.
In view of this distinct length scale separation,
$\ell_{b}\gg\ell_{a}$, the electron distribution in
Eq.~\eqref{f-part-eq} is established at the first stage of the
thermalization process. However, for longer wires, $L\gg\ell_{b}$,
relaxation of counterpropagating electrons becomes so efficient that
the temperatures $\mathcal{T}^{R/L}$ and boost velocities $u^{R/L}$
of right and left movers become equal,
$\mathcal{T}^R=\mathcal{T}^L=\mathcal{T}$ and $u^R=u^L=u$, due to
energy and momentum exchange. At the same time, the chemical
potentials $\mu^{R/L}$ are still unequilibrated in this regime,
$\Delta\mu=\mu^R-\mu^L\neq0$, since the numbers of right- and
left-moving electrons are still independently conserved. As a
result, for wires with length $L\gg\ell_b$ the distribution function
\eqref{f-part-eq} transforms into
\begin{equation}\label{f-part-eq-1}
f_p
=\frac{\theta(p)}{e^{\frac{\varepsilon_p-pu-\mu^R}{\mathcal{T}}}+1}+
\frac{\theta(-p)}{e^{\frac{\varepsilon_p-pu-\mu^L}{\mathcal{T}}}+1}\,.
\end{equation}
Because $\Delta\mu\neq0$ we refer to the states of electron system described
by the distributions in Eqs.~\eqref{f-part-eq} and \eqref{f-part-eq-1} as the
states of \textit{partial} equilibration.

\subsection{Fully equilibrated electrons}

Energy and momentum conservation allow for the scattering process in
which an electron at the bottom of the band is backscattered by two
other particles near the Fermi level [see Fig.~\ref{Fig2}(c)]. This
is the basic three-particle process that changes the numbers of
right and left movers before and after collision. In particular, the
exponentially small discontinuity of the distributions
Eqs.~\eqref{f-part-eq} and \eqref{f-part-eq-1} at $p=0$ will be
smeared by collisions of this type.

Complete equilibration of electrons, namely, relaxation of
$\Delta\mu$, relies on the electron backscattering from the right to
the left Fermi point. One should notice here that it is impossible
to realize such scattering directly since it requires momentum
transfer of $2p_F$ while Fermi blocking restricts typical momentum
exchange in the collision to $\delta p\sim T/v_F\ll p_F$. As a
consequence, complete electron backscattering, and thus relaxation
of the chemical potential difference $\Delta\mu$, occurs via a large
number of small steps $\delta p$ in momentum space such that $\delta
p\ll p_F$.~\cite{JTK} In its passage between the subsystems of right
and left movers the backscattered electron has to pass the
bottleneck of occupied states at the bottom of the band [see
Fig.~\ref{Fig2}(d)]. As a result, backscattering of electrons is
exponentially suppressed by the probability $\sim e^{-\mu/T}$ to
find an unoccupied state at the bottom of the band, and thus the
equilibration length $\ell_{eq}$ for the relaxation of the
difference in chemical potentials of left and right movers is
exponentially large, $\ell_{eq}\propto e^{\mu/T}$. For sufficiently
long wires, $L\gg\ell_{eq}$, the state of \textit{full}
equilibration is achieved and described by the
distribution~\cite{JTK,TJK}
\begin{equation}\label{f-eq}
f_p=\frac{1}{e^{\frac{\varepsilon_p-v_dp-\mu_{eq}}{\mathcal{T}}}+1}\,,
\end{equation}
where the chemical potential $\mu_{eq}$ inside the equilibrated wire
is, in general, different from $\mu_{l(r)}$ in the leads. In
Eq.~\eqref{f-eq} $v_d=I/ne$ is the electron drift velocity, where
$I$ is the electric current and $n$ is the electron density. The
partially equilibrated distribution function given by
Eq.~\eqref{f-part-eq-1} smoothly interpolates to the state of full
equilibration, Eq.~\eqref{f-eq}, when the length of the wire exceeds
the equilibration length $\ell_{eq}$. The fully equilibrated
distribution Eq.~\eqref{f-eq} is obtained from
Eq.~\eqref{f-part-eq-1} by setting $\mu^R=\mu^L=\mu_{eq}$; thus
$\Delta\mu=0$, and also $u=v_d$.

\subsection{Brief summary}

The regime of \textit{partial} equilibration described by the
distribution function in Eq.~\eqref{f-part-eq-1} covers a wide range
of lengths, $\ell_a\ll L\lesssim\ell_{eq}$.  It is more likely to be
realized in experiments than the fully equilibrated regime
\eqref{f-eq}, as the length scale $\ell_{eq}$ is exponentially
large.  Depending on the wire length $L$ a particular state of the
electron system is characterized by the extent to which the
difference in chemical potentials $\Delta\mu$ and temperatures
$\Delta T$ of left and right movers has relaxed. The recent works
Refs.~\onlinecite{TJK,ATJK} addressed transport properties of wires
with lengths in the range
\begin{equation}
\ell_a\ll\ell_b\ll L\sim\ell_{eq},
\end{equation}
which covers the crossover from the partially equilibrated regime
Eq.~\eqref{f-part-eq-1} to the fully equilibrated regime
Eq.~\eqref{f-eq}. The major emphasis of these works was on the
effect of equilibration due to electron backscattering
[Fig.~\ref{Fig2}(d)]. The main focus of the present paper is on the
transport properties of partially equilibrated wires with length
\begin{equation}
\ell_a\ll L\sim\ell_{b}\ll\ell_{eq}.
\end{equation}
In this regime the numbers $N^L$ and $N^R$ of the left- and
right-moving electrons are individually conserved up to corrections
as small as $e^{-\mu/T}$, so that electrons with energies near the
Fermi level pass through the wire without backscattering
[Fig.~\ref{Fig2}(c)]. This automatically implies that the
conductance $G$ unchaged intact by interactions and is still given
by Eq.~\eqref{G-0}. However, electrons will experience other
multiple three-particle collisions [Fig.~\ref{Fig2}(b)], which allow
momentum and energy exchange within and between the two branches of
the spectrum, thus altering thermal transport properties. The role
of these processes was not explored in previous studies devoted to
the transport in partially equilibrated quantum
wires.~\cite{TJK,ATJK}

\section{Boltzmann equation formalism}\label{Sec-BE}
\subsection{Three-particle collision integral}

Consider a quantum wire of length $L$, connected by ideal
reflectionless contacts to noninteracting leads which are biased by
a temperature difference $\Delta T$ (see Fig.~\ref{Fig1}). In the
following, we are interested only in the thermal transport
properties of the wire and assume that there is no external voltage
bias, $V=0$. We describe weakly interacting one-dimensional
electrons in the framework of the Boltzmann kinetic equation
\begin{equation}\label{BE}
v_p\partial_x f(p,x)=\mathcal{I}\{f(p,x)\}\,,
\end{equation}
where $v_p=p/m$ is the electron velocity and the evolution of the
distribution function is governed by the collision integral
$\mathcal{I}\{f(p,x)\}$. We consider the steady-state setup in which
the distribution function does not depend explicitly on time. The
collision integral, in general, is a nonlinear functional of
$f(p,x)$, whose form is determined by the scattering processes
affecting the distribution function. As discussed above, in our case
the dominant processes are three-particle collisions. Assuming that
the collision integral is local in space, we have
\begin{eqnarray}\label{Coll-Int}
\mathcal{I}\{f_1\}&=&-\hskip-.4cm\sum_{p_2,p_3\atop
p_{1'},p_{2'},p_{3'}}\!\!\! W^{1'2'3'}_{123}
\nonumber\\
&& \times\left[f_1f_2f_3(1-f_{1'})
(1-f_{2'})(1-f_{3'})\right.\nonumber\\
&& \left.-f_{1'} f_{2'}f_{3'}(1-f_1)(1-f_2)(1-f_3)\right],
\end{eqnarray}
where $W^{1'2'3'}_{123}$ is the scattering rate from the incoming
states $\{p_1,p_2,p_3\}$ into the outgoing states
$\{p_{1'},p_{2'},p_{3'}\}$, and we used the shorthand notation
$f_i=f(p_i,x)$. The Boltzmann equation [Eq.~\eqref{BE}] is
supplemented by the boundary conditions stating that the
distribution $f(p,0)$ of right-moving electrons ($p>0$) at the left
end of the wire and $f(p,L)$ of left-moving electrons ($p<0$) at the
right end coincide with the distribution function in the leads,
Eq.~\eqref{f-0}. We note here that although Eq.~\eqref{Coll-Int} is
written for the spinless case our subsequent analysis and solution
of the Boltzmann equation presented in
Sec.~\ref{Sec-BE-A}--\ref{Sec-BE-D} is applicable to the spinful
electrons as well.

An exact analytical solution of the Boltzmann equation
[Eq.~\eqref{BE}] is, in general, very difficult to find due to the
nonlinearity of the collision integral Eq.~\eqref{Coll-Int}. A
simplification is, however, possible in the case of a linear
response analysis in the externally applied perturbation (in our
case, the temperature difference $\Delta T$). Then the collision
integral can be linearized near its unperturbed value. It is
convenient to present $f(p,x)$ as
\begin{equation}\label{f-linearization}
f(p,x)=f^0_p+f^0_p(1-f^0_p)\psi(p,x)\,,
\end{equation}
where $f^0_p=(e^{(\varepsilon_p-\mu)/T}+1)^{-1}$ is the equilibrium
Fermi distribution function and $\psi(p,x)\propto\Delta T$. When
linearizing Eq.~\eqref{Coll-Int} with respect to $\psi(p,x)$ the
factor $f^0_p(1-f^0_p)$ in Eq.~\eqref{f-linearization} makes it
convenient to use the detailed balance condition
\begin{eqnarray}\label{det-balance}
f^0_{p_1}f^0_{p_2}f^0_{p_3}(1-f^0_{p_{1'}})
(1-f^0_{p_{2'}})(1-f^0_{p_{3'}})=\nonumber\\
f^0_{p_{1'}}
f^0_{p_{2'}}f^0_{p_{3'}}(1-f^0_{p_1})(1-f^0_{p_2})(1-f^0_{p_3})\,,
\end{eqnarray}
valid at $\varepsilon_{p_1}+\varepsilon_{p_2}+\varepsilon_{p_3}
=\varepsilon_{p_{1'}}+\varepsilon_{p_{2'}}+\varepsilon_{p_{3'}}$.
Substituting Eq.~\eqref{f-linearization} into the collision integral
and using Eq.~\eqref{det-balance} one arrives at the linearized
version of Eq.~\eqref{Coll-Int}
\begin{eqnarray}\label{Coll-Int-Linear}
\!\!\!\!\!\mathcal{I}\{\psi(p_1,x)\}\!&=&\!-\!\!\!\!\!\sum_{p_2,p_3\atop
p_{1'},p_{2'},p_{3'}}\!\!\!
\mathbb{K}^{1'2'3'}_{123}\nonumber\\
&&\!\!\!\times\left[\psi(p_1,x)+\psi(p_2,x)+\psi(p_3,x)\right.\nonumber\\
&&\!\!\!\left.-\psi(p_{1'},x)-\psi(p_{2'},x)-\psi(p_{3'},x)\right]\,,
\end{eqnarray}
with the kernel $\mathbb{K}$ defined as
\begin{eqnarray}\label{Delta-kernel}
\mathbb{K}^{1'2'3'}_{123}&=&W^{1'2'3'}_{123}f^0_{p_1}f^0_{p_2}f^0_{p_3}\nonumber\\
&&\times (1-f^0_{p_{1'}})(1-f^0_{p_{2'}})(1-f^0_{p_{3'}})\,.
\end{eqnarray}
The explicit form of the scattering rate $W^{1'2'3'}_{123}$ is not
important for the following discussion. It is discussed in detail
in Appendix~\ref{Sec-Appendix-A}.

\subsection{Solution strategy}\label{Sec-BE-A}

Even after linearization the solution of the integral Boltzmann
equation that satisfies given boundary conditions is still a
complicated problem. However, our task is simplified greatly since
we already know the structure of the distribution function $f(p,x)$.
Indeed, we have discussed in Sec.~\ref{Sec-Preview} that for wires
with length $\ell_a\ll L\sim\ell_b$ the electron system is in the
regime of partial equilibration with the distribution function given
by Eq.~\eqref{f-part-eq}. Thus, the class of functions we need to
consider to solve our boundary problem is, in fact, rather narrow.
The solution we are seeking is conveniently parametrized by six
unknowns: $\mu^{R(L)}(x)$, $u^{R(L)}(x)$, and
$\mathcal{T}^{R(L)}(x)$. Instead of solving one integro-differential
equation for $f(p,x)$ we will reduce our task to solving a system of
six linear ordinary differential equations that govern the spatial
evolution of the parameters defining the distribution function in
Eq.~\eqref{f-part-eq}. This is possible since the momentum
dependence of the distribution function is fully determined by our
ansatz \eqref{f-part-eq}, which allows completion of all $p$
integrations in the Boltzmann equation analytically. Among the six
equations we need, four represent conservation laws: conservation of
numbers of right- and left-moving electrons $N^{R/L}$, of total
momentum $P$, and of energy $E$ of the electron system. The other
two are kinetic equations that account for the momentum and energy
exchange between subsystems of right and left movers, thus capturing
the processes of thermalization.

For the following analysis it is convenient to measure the chemical
potentials and temperatures from their equilibrium values:
$\mu^{R(L)}(x)=\mu+\delta\mu^{R(L)}(x)$ and
$\mathcal{T}^{R(L)}(x)=T+\delta \mathcal{T}^{R(L)}(x)$. Expanding
now Eq.~\eqref{f-part-eq} to the linear order in $u^{R(L)}$,
$\delta\mu^{R(L)}$ and $\delta \mathcal{T}^{R(L)}$ and using
Eq.~\eqref{f-linearization} we can identify $\psi(p,x)$ that enters
the collision integral in Eq.~\eqref{Coll-Int-Linear} as
\begin{equation}\label{psi}
\psi(p,x)=\psi^R(p,x)+\psi^L(p,x)
\end{equation}
where
\begin{eqnarray}\label{psi-RL}
&&\hskip-.55cm\psi^{R(L)}(p,x)=\theta(\pm
p)\nonumber\\
&&\hskip-.55cm\times
\left[\psi^{R(L)}_{\mu}(p,x)+\psi^{R(L)}_{u}(p,x)+\psi^{R(L)}_{\mathcal{T}}(p,x)\right].
\end{eqnarray}
Here the three contributions are
\begin{eqnarray}
&&\psi^{R(L)}_{\mu}(p,x)=\frac{\delta\mu^{R(L)}(x)}{T}\,,\label{psi-RL-mu}\\
&&\psi^{R(L)}_u(p,x)=\frac{pu^{R(L)}(x)}{T}\,,\label{psi-RL-u}\\
&&\psi^{R(L)}_{\mathcal{T}}(p,x)=\frac{(\varepsilon_p-\mu)\delta
\mathcal{T}^{R(L)}(x)}{T^2}\,.\label{psi-RL-t}
\end{eqnarray}
These functions evolve in the real space as prescribed by the
collision integral Eq.~\eqref{Coll-Int-Linear} while their boundary
values can be extracted from the respective distributions in the
leads [Eq.~\eqref{f-0}]. Indeed, expanding Eq.~\eqref{f-0} with
$V=0$ one obtains
\begin{equation}
f_p=f^0_p+\frac{(\varepsilon_p-\mu)\Delta
T}{2T^2}f^0_p(1-f^0_p)[\theta(p)-\theta(-p)].
\end{equation}
By matching this result to Eq.~\eqref{f-linearization} with
$\psi(p,x)$ taken from Eqs.~\eqref{psi} and \eqref{psi-RL} we deduce
the boundary conditions
\begin{eqnarray}\label{boundary-cond}
&&\delta\mu^R(0)=\delta\mu^L(L)=0\,,\label{boundary-cond-1}\\
&&u^{R}(0)=u^L(L)=0\,,\label{boundary-cond-2}\\
&&\delta\mathcal{T}^R(0)=-\delta\mathcal{T}^L(L)=\Delta
T/2\,.\label{boundary-cond-3}
\end{eqnarray}
Our task now is to derive the set of coupled ordinary differential
equations that govern the spatial evolution of the unknown
parameters $u^{R(L)}(x)$, $\delta\mu^{R(L)}(x)$, and
$\delta\mathcal{T}^{R(L)}(x)$. This will give us complete knowledge
of the electron distribution function. Knowing all parameters in
Eq.~\eqref{f-part-eq} we will be able to find the heat current and
finally the thermal conductance of the system. Before we realize
this plan, the conservation laws must be discussed.

\subsection{Transport currents and conservation laws}

Conservation of the total number of particles implies that in a
steady state the particle current $I(x)$ is uniform along the wire.
Correspondingly, we infer from the conservation of total momentum
$P$ and total energy $E$ that in the steady state a constant
momentum current $I_P$ and a constant energy current $I_E$ flow
through the system. In the following it will be convenient to
express these currents as the sums of individual contributions of
the left- and right-moving electrons, e.g., $I= I^R+ I^L$, thus
introducing:
\begin{eqnarray}
&&\hskip-.5cm I^{R(L)}(x)=\int^{+\infty}_{-\infty}\frac{dp}{h}\,
\theta(\pm p)v_pf(p,x)\,,\label{I}\\
&&\hskip-.5cm I^{R(L)}_P(x)=\int^{+\infty}_{-\infty}\frac{dp}{h}\,
\theta(\pm p)pv_pf(p,x)\,,\label{I-P}\\
&&\hskip-.5cm
I^{R(L)}_E(x)=\int^{+\infty}_{-\infty}\frac{dp}{h}\,\theta(\pm
p)\varepsilon_pv_pf(p,x)\label{I-E}\,.
\end{eqnarray}
The positive sign in the step function corresponds to right movers
and the negative one to left movers. Since we neglect small
backscattering effects, the numbers of right- and left-moving
electrons are conserved independently:
\begin{equation}
\dot{N}^{R/L}=0\,.
\end{equation}
It follows then immediately from the continuity equations that
particle currents are uniform along the wire,
\begin{equation}\label{Conserv-I}
\partial_x I^R(x)=0\,,\qquad \partial_x I^{L}(x)=0\,.
\end{equation}
Similarly we present the conservation of total momentum and energy,
\begin{eqnarray}
&&\partial_x \big[I^R_P(x)+I^L_P(x)\big]=0\,,\label{Conserv-I-P}\\
&&\partial_x \big[I^R_E(x)+I^L_E(x)\big]=0\,.\label{Conserv-I-E}
\end{eqnarray}
As the next step we express the currents \eqref{I}--\eqref{I-E} in
terms of the parameters defining the electron distribution function.
Specifically, we use Eq.~\eqref{f-linearization} and
Eqs.~\eqref{psi-RL}--\eqref{psi-RL-t} together with the current
definition in Eq.~\eqref{I} and thus find from Eq.~\eqref{Conserv-I}
\begin{equation}\label{mu-eq}
\frac{d\delta\mu^{R(L)}}{dx}=\mp p_F\frac{du^{R(L)}}{dx}
\left(1-\frac{\pi^2T^2}{24\mu^2}-\frac{7\pi^4T^4}{384\mu^4}\right)\,.
\end{equation}
When deriving this equation from Eq.~\eqref{I} we had to carry out a
Sommerfeld expansion up to the fourth order in $T/\mu\ll1$. Note
here that even though backscattering is neglected $\delta\mu^{R(L)}$
must change in space to accommodate the conservation laws for
currents [Eqs.~\eqref{Conserv-I}--\eqref{Conserv-I-E}], and
$\mu^{R(L)}(x)=\mu+\delta\mu^{R(L)}(x)$ coincide with
$\mu_{r(l)}=\mu$ only at the ends of the wire [see the boundary
conditions Eq.~\eqref{boundary-cond-1}].

We then perform a similar calculation for the momentum and energy
currents. From the momentum conservation Eq.~\eqref{Conserv-I-P}, we
find
\begin{eqnarray}\label{Eq-Ip}
p_F\frac{T}{2\mu}\left(1+\frac{5\pi^2}{12}\frac{T^2}{\mu^2}\right)\left(\frac{d
u^R}{dx}-\frac{du^L}{dx}\right)+\nonumber\\
\left(1+\frac{7\pi^2}{40}\frac{T^2}{\mu^2}\right)\left(\frac{d\delta
\mathcal{T}^R}{dx}+\frac{d\delta\mathcal{T}^L}{dx}\right)=0\,,
\end{eqnarray}
while from the energy conservation, Eq.~\eqref{Conserv-I-E},
\begin{eqnarray}\label{Eq-Ie}
p_F\frac{T}{2\mu}\left(1+\frac{7\pi^2}{40}\frac{T^2}{\mu^2}\right)\left(\frac{d
u^R}{dx}+\frac{du^L}{dx}\right)+\nonumber\\
\left(\frac{d\delta\mathcal{T}^R}{dx}-\frac{d\delta
\mathcal{T}^L}{dx}\right)=0\,.
\end{eqnarray}
When deriving these two equations we also made use of
Eq.~\eqref{mu-eq} to exclude the chemical potentials of the right
and left movers.

\subsection{Scattering processes and kinetic equations}\label{Sec-BE-D}

Although the total momentum and energy currents $I_P$ and $I_E$ are
conserved, such currents taken for the left and right movers
separately, $I^{R/L}_{P}$ and $I^{R/L}_{E}$, are not. Indeed, the
three-particle collisions shown in Fig.~\ref{Fig2}(b) induce
momentum and energy exchange between counter-propagating electrons.
Let us focus on a small segment of the wire between the positions
$x$ and $x+\Delta x$, where $0<x<L$. The difference
$I^{R/L}_P(x+\Delta x)-I^{R/L}_{P}(x)$ is equal to the rate of
change of the momentum of right-moving electrons
$\dot{P}^{R/L}=\dot{p}^{R/L}\Delta x$. Here $\dot{p}^{R/L}$ is the
rate per unit of length. As a result, the continuity equation for
the momentum exchange reads
\begin{equation}\label{Eq-Ip-Kin}
\partial_x \big[I^R_P(x)-I^L_P(x)\big]=2\dot{p}^R\,.
\end{equation}
Here we used $\dot{p}^L=-\dot{p}^R$, which is ensured by the
conservation of total momentum. In complete analogy we can now
relate the difference of energy currents $I^R_E(x+\Delta
x)-I^R_{E}(x)$ to the corresponding energy exchange rate
$\dot{E}^R=\dot{e}^R\Delta x$, which gives us
\begin{equation}\label{Eq-Ie-Kin}
\partial_x \big[I^R_E(x)-I^R_E(x)\big]=2\dot{e}^R\,,
\end{equation}
where we also used $\dot{e}^L=-\dot{e}^R$ guaranteed by the energy
conservation. The right-hand side of Eqs.~\eqref{Eq-Ip-Kin} and
\eqref{Eq-Ie-Kin} can be calculated from the collision integral of
the Boltzmann equation [Eq.~\eqref{Coll-Int-Linear}].

There are two basic processes which contribute to $\dot{p}^R$ and
$\dot{e}^R$. The first one includes two right movers that scatter
off one left mover. This process is shown in Fig.~\ref{Fig2}(d). The
other process, when two left movers scatter off one right mover, is
equally important. Keeping both terms, using
Eqs.~\eqref{Coll-Int-Linear} and \eqref{psi} we find for the
relaxation rates (details of the derivation are given in Appendix
\ref{Sec-Appendix-PE})
\begin{eqnarray}
&&\hskip-.7cm\dot{p}^R=-k_F\frac{p_F}{2\tau}\frac{u^R-u^L}{v_F}\,,\label{PR}\\
&&\hskip-.7cm\dot{e}^R=-k_F\frac{\mu}{\tau}\frac{\delta
\mathcal{T}^R-\delta
\mathcal{T}^L}{T}\,,\label{ER}\\
&&\hskip-.7cm\frac{1}{\tau}=\frac{3}{k_F\Delta
x}\sum_{p_1>0,p_2>0,p_3<0\atop
p_{1'}>0,p_{2'}>0,p_{3'}<0}\frac{v^2_F(p_{3'}-p_3)^2}{\mu
T}\mathbb{K}^{1'2'3'}_{123}\label{tau-E}.
\end{eqnarray}
Having determined relaxation the rates we now return to
Eqs.~\eqref{Eq-Ip-Kin} and \eqref{Eq-Ie-Kin}. Computing the momentum
and energy currents of right and left movers in the same way as we
did in the previous section from Eqs.~\eqref{I-P}--\eqref{I-E} and
using the relaxation rates from Eqs.~\eqref{PR}--\eqref{ER} we find
two additional equations which describe the relaxation of momentum:
\begin{eqnarray}\label{Eq-Ip-Kin-1}
&& p_F\frac{\pi^2T^2}{12\mu^2}\!
\left(\frac{du^R}{dx}+\frac{du^L}{dx}\right)\!\!
\left(1+\frac{5\pi^2}{12}\frac{T^2}{\mu^2}\right)\nonumber\\
&&+
\frac{\pi^2T}{6\mu}\!\left(\frac{d\delta\mathcal{T}^R}{dx}-\frac{d\delta
\mathcal{T}^R}{dx}\right)\!\!\left(1+\frac{7\pi^2}{40}\frac{T^2}{\mu^2}\right)\nonumber\\
&&=\!-\frac{hk_F}{\tau}\frac{u^R-u^L}{v_F}
\end{eqnarray}
and energy
\begin{eqnarray}\label{Eq-Ie-Kin-1}
&&\hskip-.5cm
p_F\frac{\pi^2T^2}{6\mu^2}\left(\frac{du^R}{dx}-\frac{du^L}{dx}\right)
\!\!\left(1+\frac{7\pi^2}{40}\frac{T^2}{\mu^2}\right)+\nonumber\\
&&\hskip-.5cm \frac{\pi^2T}{3\mu}\!\left(\frac{d\delta
\mathcal{T}^R}{dx}+\frac{d\delta
\mathcal{T}^L}{dx}\right)\!=-\frac{2hk_F}{\tau}\frac{\delta
\mathcal{T}^R-\delta \mathcal{T}^L}{T}.
\end{eqnarray}
Equations \eqref{mu-eq}--\eqref{Eq-Ie} and
\eqref{Eq-Ip-Kin-1}--\eqref{Eq-Ie-Kin-1}, together with the boundary
conditions Eqs.~\eqref{boundary-cond-1}--\eqref{boundary-cond-3},
represent the closed system of six coupled differential equations
whose solution fully determines the six parameters $\mu^{R/L},
u^{R/L}, \mathcal{T}^{R/L}$ that define the electron distribution
function~\eqref{f-part-eq}. We now find these parameters explicitly.
For that purpose let us introduce dimensionless variables
\begin{equation}\label{dim-units}
\eta_\pm=\frac{u^{R}\pm u^L}{v_F}\,,\qquad \theta_\pm=\frac{\delta
\mathcal{T}^{R}\pm\delta \mathcal{T}^L}{T}\,,
\end{equation}
and the microscopic scattering length
\begin{equation}\label{l-Q-def}
\ell_b=\frac{\pi^3T^4}{360\mu^4}(v_F\tau)\,.
\end{equation}
Calculation of $\ell_b$ requires a detailed knowledge of the
scattering rate, implicit in the kernel $\mathbb{K}^{1'2'3'}_{123}$,
as a function of momenta transferred in a collision. In Appendix
\ref{Sec-Appendix-A} we provide this information for the case of the
three-particle collisions under consideration and in Appendix
\ref{Sec-Appendix-l-Q} find $\ell_b$ explicitly for the spinless and
spinful cases.

After some algebra the coupled equations \eqref{Eq-Ip},
\eqref{Eq-Ie} and \eqref{Eq-Ip-Kin-1}, \eqref{Eq-Ie-Kin-1} can be
reduced to the following form:
\begin{eqnarray}
&&\frac{\partial \theta_+}{\partial x}=-\alpha \frac{\partial\eta_-}{\partial x}\,,\label{1}\\
&&\frac{\partial\eta_+}{\partial x}=-\beta \frac{\partial\theta_-}{\partial x}\,,\label{2}\\
&&\frac{\partial\theta_-}{\partial x}=\frac{\eta_-}{\ell_b}\,,\label{3}\\
&&\frac{\partial\eta_-}{\partial
x}=\frac{\theta_-}{\ell_b}\,,\label{4}
\end{eqnarray}
which contain two dimensionless parameters
\begin{eqnarray}\label{a-b}
\alpha=1+\frac{29\pi^2}{120}\frac{T^2}{\mu^2}\,,
\quad\beta=1-\frac{7\pi^2}{40}\frac{T^2}{\mu^2}\,,
\end{eqnarray}
where corrections of higher order in $T/\mu\ll1$ were neglected. The
remaining two equations for the chemical potentials of right and
left movers [Eq.~\eqref{mu-eq}] are not written here for brevity.
The latter do not enter the heat current and thus are not explicitly
needed. Equations \eqref{1}--\eqref{4} can now be easily solved,
with the result
\begin{equation}\label{tR}
\theta^R(x)=\frac{\Delta
T}{2T}\frac{\alpha_-e^{x/\ell_b}+\alpha_+e^{(L-x)/\ell_b}}
{\alpha_-+\alpha_+e^{L/\ell_b}}\,,
\end{equation}
\begin{equation}\label{tL}
\theta^L(x)=-\frac{\Delta
T}{2T}\frac{\alpha_+e^{x/\ell_b}+\alpha_-e^{(L-x)/\ell_b}}
{\alpha_-+\alpha_+e^{L/\ell_b}}\,,
\end{equation}
\begin{equation}\label{uR}
\eta^R(x)=\frac{\Delta
T}{2T}\frac{\beta_-(e^{x/\ell_b}-1)-\beta_+(e^{(L-x)/\ell_b}-e^{L/\ell_b})}
{\alpha_-+\alpha_+e^{L/\ell_b}}\,,
\end{equation}
\begin{equation}\label{uL}
\eta^L(x)=-\frac{\Delta
T}{2T}\frac{\beta_+(e^{x/\ell_b}-e^{L/\ell_b})-\beta_-(e^{(L-x)/\ell_b}-1)}
{\alpha_-+\alpha_+e^{L/\ell_b}}\,,
\end{equation}
where $\alpha_\pm=1\pm\alpha$, $\beta_\pm=1\pm\beta$ and
$\theta^{R/L}=(\theta_+\pm\theta_-)/2$,
$\eta^{R/L}=(\eta_+\pm\eta_-)/2$. This concludes our solution of the
Boltzmann equation for three-particle collisions.

\section{Heat current and thermal conductance}\label{Sec-Summary}

Complete knowledge of the distribution function \eqref{f-part-eq}
allows us to compute physical observables. Specifically, we are
interested in the thermal conductance $K$. For the latter we need to
evaluate the heat current
\begin{equation}\label{I-Q}
I_Q(x)=I_E(x)-\mu I(x)\,.
\end{equation}
By using Eqs.~\eqref{f-part-eq} we carry out a Sommerfeld expansion
for the particle and energy currents $I$ and $I_E$ from
Eqs.~\eqref{I} and \eqref{I-E}, up to the fourth order in
$T/\mu\ll1$, and then find from the above definition
[Eq.~\eqref{I-Q}]
\begin{equation}
I_Q(x)=\frac{\pi^2T^2}{3h}\left[\frac{\eta_+(x)}{\beta}+\theta_-(x)\right],
\end{equation}
which is presented here in our notation defined in
Eq.~\eqref{dim-units}. With the help of Eqs.~\eqref{tR}--\eqref{uL}
it can be readily checked that $I_Q$ is uniform along the wire. This
fact is \textit{a priori} expected and follows from the conservation
laws, which we already explored above. By knowing $I_Q$ we can
finally find the thermal conductance $K(L)=I_Q/\Delta T$ as a
function of the wire length,
\begin{eqnarray}\label{K(L)}
\frac{K(L)}{K_0}=\frac{\tanh(L/2\ell_b)+\beta}{\alpha\beta\tanh(L/2\ell_b)+\beta}\,.
\end{eqnarray}
This is the main result of our paper. Note here that $K_0=\pi^2T/3h$
for the case of spinless electrons, whereas $K_0$ is given by
Eq.~\eqref{K-0} for electrons with spin. The functional form of
$K(L)$ remains the same in both cases except for the expressions for
$\ell_b$, which we discuss below. Let us now analyze limiting cases
of Eq.~\eqref{K(L)} and discuss the microscopic form of the
scattering length $\ell_b$.

Equation \eqref{K(L)} interpolates smoothly between two distinct
limits. In short wires, $L\ll\ell_b$, from the expansion of
Eq.~\eqref{K(L)} one obtains for the interaction-induced correction
to thermal conductance, $\delta K=K-K_0$, the following result:
\begin{equation}\label{delta-K-1}
\frac{\delta K(L)}{K_0}= -
\frac{\pi^2}{30}\frac{T^2}{\mu^2}\frac{L}{\ell_{b}}\,,\quad
L\ll\ell_b\,.
\end{equation}
In such short wires electrons propagate from one lead to the other,
rarely experiencing three-particle collisions of the type shown in
Fig.~\ref{Fig2}(b). Thus their distribution function is
approximately determined by that in the leads [Eq.~\eqref{f-0}].
Under this assumption one can adopt the strategy of
Ref.~\onlinecite{Lunde}, applied previously for the calculation of
conductance and thermopower in short wires, and treat the collision
integral of the Boltzmann equation perturbatively, thus neglecting
effects of thermalization on the distribution function. Technically
speaking, this corresponds to a lowest order iteration for the
Boltzmann equation, which amounts to substituting distribution
\eqref{f-0} into the collision integral \eqref{Coll-Int-Linear} to
calculate the correction to $I_Q$. This perturbative procedure
immediately reproduces Eq.~\eqref{delta-K-1}.

\begin{figure}
  \includegraphics[width=8cm]{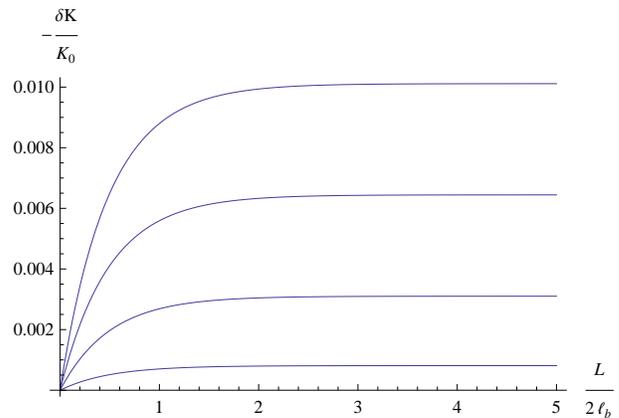}
  \caption{(Color online) Interaction-induced correction to the thermal conductance of a clean
  quantum wire as a function of its length plotted for different values of temperature
  (from the bottom to the top curve): $T/\mu=0.05, 0.1, 0.15, 0.2$. For $L\ll\ell_{b}$ the correction
  scales with $L$ and saturates to a constant value $\propto \left(T/\mu\right)^2$
  once $\ell_{b}\ll L$ in accordance with Eqs.~\eqref{delta-K-1} and \eqref{delta-K-2}.
  }\label{Fig3}
\end{figure}

It is physically expected that in longer wires particle collisions
Fig.~\ref{Fig2}(b) should have a much more dramatic effect on the
distribution function and thus thermal transport. Indeed, once full
thermalization has been achieved for $L\gg\ell_b$ we find from
Eq.~\eqref{K(L)} that the correction to thermal conductance
saturates:
\begin{equation}\label{delta-K-2}
\frac{\delta K(L)}{K_0} = -\frac{\pi^2}{30}\frac{T^2}{\mu^2}\,,\quad
\ell_b\ll L\ll\ell_{eq}\,.
\end{equation}
One interesting aspect of Eq.~\eqref{delta-K-2} is that $\delta K$
is independent of the interaction strength. It means that no matter
how weak the interactions are, for sufficiently long wires
thermalization between right- and left-moving electrons is
eventually established, which leads to saturation of $\delta K$. The
behavior of $\delta K(L)$ as a function of the wire length is
summarized Fig.~\ref{Fig3}.

The interaction strength, however, sets the length scale $\ell_b$ at
which thermalization occurs. Its actual dependence on temperature is
determined by the phase space available for a three-particle
collision to occur and by the dependence of the corresponding
scattering amplitude on momenta transferred in a collision. For
spinless electrons and Coulomb interaction we find (see Appendix
\ref{Sec-Appendix-l-Q} for the derivation and additional
discussions)
\begin{equation}\label{l-Q-spinless}
\ell^{-1}_{b}\simeq k_F\lambda_1(k_Fw)(e^2/\hbar
v_F\kappa)^4(T/\mu)^3\,,
\end{equation}
where $w$ is thw wire width and $\lambda_1(z)=z^4\ln^2(1/z)$. In the
case of spinful electrons, the scattering length changes to
\begin{equation}\label{l-Q-spinfull}
\ell^{-1}_{b}\simeq k_F\lambda_2(k_Fw)(e^2/\hbar
v_F\kappa)^4(T/\mu)\ln^2(\mu/T)\,,
\end{equation}
where $\lambda_2(z)=\ln^2(1/z)$. Contrasting
Eqs.~\eqref{l-Q-spinless} and \eqref{l-Q-spinfull}, one sees that
the spin of the electron plays an important role since the inverse
scattering length of spinful electrons is significantly larger, by a
factor of $(\mu/T)^2\gg1$. This is a manifestation of the Pauli
exclusion principle. Indeed, for three-particle scattering to occur
electrons must approach each other on a distance of the order of
$\sim k^{-1}_{F}$. When electrons are spinless the Pauli exclusion
suppresses the probability of such scattering. In contrast, the
suppression is not as strong when the total spin of the three
colliding particles is $1/2$ since at least two electrons may have
opposite spins while exclusion applies to the third particle. This
technical point and the importance of the exchange effect in the
scattering amplitudes are discussed in more detail in Appendix
\ref{Sec-Appendix-A}.

\section{Discussion}\label{Sec-Discussion}

In this paper we studied the thermal transport properties of
one-dimensional electrons in quantum wires. In this system
equilibration is strongly restricted by the phase space available
for electron scattering and conservation laws such that leading
effects stem from the three-particle collisions. This is in sharp
contrast with higher-dimensional systems where already pair
collisions provide electronic relaxation. Although our theory is
applicable only in the weakly interacting limit, the results
presented are still beyond the picture of the Luttinger liquid since
three-particle collisions are not captured by the latter. We have
elucidated the microscopic processes involved in electron
thermalization and developed a scheme for solving the Boltzmann
equation analytically within a linear response analysis. Our
approach allows us to find the thermal conductance for arbitrary
relations between the wire length and microscopic relaxation length
[see Eq.~\eqref{K(L)}].

In order to establish a connection to previous work~\cite{TJK,ATJK}
we emphasize that our solution of the kinetic equations and the
result for thermal conductance presented in Eq.~\eqref{K(L)} rely on
the simplifying assumption that electron backscattering can be
neglected. This is a good approximation except for the case of very
long wires, $L\gtrsim\ell_{eq}$, where the small probability of
backscattering $\sim e^{-\mu/T}$ is compensated by the large phase
space available for scattering to happen. Accounting for the
backscattering processes, it was found in Ref.~\onlinecite{TJK} that
for wires with length $L\sim\ell_{eq}$ the thermal conductance is
\begin{equation}\label{K(L)-TJK}
\frac{K(L)}{K_0}=\frac{\ell_{eq}}{L+\ell_{eq}}\,.
\end{equation}
This result gives only an exponentially small correction to the
thermal conductance, $\delta K/K_0=-L/\ell_{eq}\propto e^{-\mu/T}$,
in the limit $L\ll\ell_{eq}$, since in the analysis of
Ref.~\onlinecite{TJK} thermalization effects on the distribution
function were neglected. It is our result Eq.~\eqref{delta-K-2} that
gives the leading-order correction to $\delta K$ in this case. On
the other hand, our expression \eqref{K(L)} is not applicable for
the long wires, $L\sim\ell_{eq}$, whereas Eq.~\eqref{K(L)-TJK} works
in this regime. It displays an essentially different feature, which
is solely due to backscattering processes, namely, the vanishing
thermal conductance $\delta K\propto 1/L$ as $L\to\infty$.

Our work may be relevant for a number of recent experiments. In
particular, the authors of Ref.~\onlinecite{Chiatti} reported
thermal conductance measurements and a lower value of $K$ than that
predicted by the Wiedemann-Franz law, at the plateau of electrical
conductance. As we explained, corrections to $G$ are exponentially
small, $G=2e^2/h-\mathcal{O}(e^{-\mu/T})$, for wires with
$L\ll\ell_{eq}$. Thus the conductance remains essentially unaffected
by interactions, and its quantization is robust. In contrast, the
effect of three-particle collisions on the thermal conductance is
much more pronounced. Our equation \eqref{delta-K-2} shows that the
thermal conductance is reduced by interactions, which is
qualitatively consistent with the experimental
observation.~\cite{Chiatti} Apparent violation of the
Wiedemann-Franz law is due to the fact that interaction-induced
corrections $\delta K$ and $\delta G$ originate from physically
distinct scattering processes [see Figs.~\ref{Fig2}(b) and
\ref{Fig2}(c), respectively].

Another experiment~\cite{Birge} reported measurements of the
electron distribution function in one-dimensional wires. This
experiment demonstrated that electrons thermalize despite the severe
constraints imposed by the conservation laws and dimensionality on
the particle collisions. We take the point of view that
three-particle collisions are responsible for relaxation and provide
an explicit solution of the Boltzmann equation, thus uncovering the
structure of the distribution function [see Eqs.~\eqref{f-part-eq}
and \eqref{tR}--\eqref{uL}], which in principle can be compared to
experimental results.~\cite{Estimate}

A related study~\cite{Barak} provided us information about the time
scales of thermalization of one-dimensional electrons. Although we
do not study the latter our results for the relaxation lengths
Eqs.~\eqref{l-Q-spinless} and \eqref{l-Q-spinfull} can be directly
linked to the experiment. Note also that the dramatic difference
between the relaxation lengths, and thus the times, of spinful and
spinless electrons provides a distinct signature of three-particle
collisions that could be tested experimentally.

There is a very important limitation on the applicability of
Eqs.~\eqref{K(L)} and \eqref{l-Q-spinfull} that we need to discuss
in the case of spinful electrons.~\cite{KOG} From the point of view
of Luttinger liquid theory, electrons are not well-defined
excitations in one dimension and instead one should use a bosonic
description in terms of charge and spin modes. The weakly
interacting limit considered here and usage of the Boltzmann
equation assumes that electrons maintain their integrity during
collisions and thus neglects effects of spin-charge separation. In
order to quantify to what extent such a description is valid,
consider an electron with excitation energy $\xi$ above the Fermi
energy $\mu$. For quadratic dispersion, $\varepsilon_p=p^2/2m$, the
velocity of such electrons differs from that of the electrons in the
Fermi sea by an amount $\Delta v=\xi/mv_F$. Spin and charge do not
separate appreciably if $\Delta v\gg v_c-v_s$, where $v_{c(s)}$ are
the velocities of charge (spin) excitations. At finite temperatures
the characteristic excitation energy is $\xi\sim T$, so that the
above condition can be equivalently reformulated as
$T/\mu\gg(v_c-v_s)/v_F$. For weakly interacting electrons the
difference between the velocities of charge and spin modes is
related to the zero-momentum Fourier component of the
electron-electron interaction potential, namely, $v_c-v_s\simeq
V_0/\pi\hbar\ll1$. This implies that at low temperatures when
$T/\mu\ll V_0/\hbar v_F$ a description in terms of electrons breaks
down and Eqs.~\eqref{K(L)} and \eqref{l-Q-spinfull} are no longer
applicable.

Finally, our work also points to open issues and directions for
future research. It is of great interest to understand the fate of
energy relaxation and the nature of thermal transport in the case of
strong interactions which simultaneously have to be combined with
nonequilibrium conditions. At very low temperatures a description of
a one-dimensional system in terms of electronic excitations becomes
inadequate even if the interactions are weak. The effect of
spin-charge separation has to be included, and thermal transport
from plasmons and their relaxation are central issues to consider.

\subsection*{Acknowledgements}

We would like to acknowledge useful discussions with A.~V.~Andreev,
N.~Andrei, N.~Birge, P.~W.~Brouwer, L.~I.~Glazman, A.~Imambekov,
A.~Kamenev, T.~Karzig, and F.~von~Oppen. This work at ANL was
supported by the U.S. DOE, Office of Science, under Contract No.
DE-AC02-06CH11357, and at ENS by the ANR Grant No. 09-BLAN-0097-01/2
(Z.R.).

\appendix
\section{Derivation of $\dot{P}^R$ and
$\dot{E}^R$}\label{Sec-Appendix-PE}

In this appendix we derive Eqs.~\eqref{PR}--\eqref{tau-E} presented
in the main text of the paper.  As explained in Sec.~\ref{Sec-BE-D},
when computing $\dot{P}^R$ and $\dot{E}^R$ we have to account for
two types of scattering process.  One is shown in Fig.~\ref{Fig2}(b)
and the other is similar and consists of a scattering of one left
mover and two right movers. We start by considering the quantity
$\mathcal{P}_n=\sum_{p_1>0} p_1^n f_1$. Its rate of change is
\begin{align}
\dot{\mathcal{P}}_n&= \sum_{p_1>0}
p_1^n\dot{f}_1=-\sum_{p_1>0,p_2,p_3
\atop p_{1'},p_{2'},p_{3'}} p_1^n\mathbb{K}^{1'2'3'}_{123}\notag\\
&\times\left(\psi_1+\psi_2+\psi_3-\psi_{1'}-\psi_{2'}-\psi_{3'}\right),
\end{align}
where we used the Boltzmann equation [Eq.~(\ref{BE})] and the
short-hand notation $\psi_i=\psi(p_i,x)$. It is convenient to split
each sum from the last equation into parts that contain positive and
negative values of the momenta, so that one gets
\begin{align}\label{Bndot}
\dot{\mathcal{P}}_n=\sum_{+ - -\atop - - -}(\ldots)+3\sum_{+ -
-\atop + + -}(\ldots)+6\sum_{+ + -\atop + - -}(\ldots)+\sum_{+ -
-\atop + + +}(\ldots)\notag\\+\sum_{+ + +\atop - -
-}(\ldots)+3\sum_{+ + +\atop + - -}(\ldots)+3\sum_{+ + +\atop + +
-}(\ldots)+2\sum_{+ + -\atop + + +}(\ldots)\notag\\+2\sum_{+ +
-\atop - - -}(\ldots)+3\sum_{+ - -\atop + - -}(\ldots)+6\sum_{+ +
-\atop + + -}(\ldots)+\sum_{+ + +\atop + + +}(\ldots).
\end{align}
The notations here are as follows
\begin{equation}
\sum_{+--\atop---}(\ldots)=\sum_{p_1>0,p_2<0,p_3<0\atop
p_{1'}<0,p_{2'}<0,p_{3'}<0}(\ldots)\,,
\end{equation}
and analogously for the other terms. When deriving Eq.~(\ref{Bndot})
we have used the following symmetry properties of the kernel: (a)
exchange of incoming and outgoing momenta
$\mathbb{K}_{123}^{1'2'3'}=\mathbb{K}_{1'2'3'}^{123}$, (b) pairwise
exchange $\mathbb{K}_{123}^{1'2'3'}=\mathbb{K}_{213}^{2'1'3'}$, and
(c) inversion of momenta $p_i\to -p_i$,
$\mathbb{K}_{123}^{1'2'3'}=\mathbb{K}_{-1-2-3}^{-1'-2'-3'}$. In the
final expression for $\dot{\mathcal{P}}_n$ we keep only the terms of
Eq.~(\ref{Bndot}) that contain equal numbers of positive incoming
and outgoing momenta, i.e., the last three terms. The other terms
contain at least one state near the bottom of the band and therefore
give a contribution that is exponentially suppressed due to the
small probability of finding an unoccupied state. After employing
Eqs.~(\ref{psi})--(\ref{psi-RL-t}) combined with momentum and energy
conservations, we end up with
\begin{eqnarray}
\dot{\mathcal{P}}_n\!\!&=\!\!& 3\sum_{++-\atop
++-}\mathbb{K}_{123}^{1'2'3'}
\bigg[\frac{\delta\mathcal{T}^R-\delta\mathcal{T}^L}{T^2}
\left[2p_1^n-(-p_3)^n\right]\notag\\
&\times&\!\!\!(\varepsilon_{p_3}-\varepsilon_{p_{3'}})
+\frac{u^L-u^R}{T}\left[2p_1^n+(-p_3)^n\right]
(p_3-p_{3'})\bigg].\nonumber
\end{eqnarray}
For $n=1$ from the last expression we easily get
\begin{align}
\dot{P}^R=\dot{\mathcal{P}}_1=-3\frac{u^R-u^L}{T}\sum_{++-\atop++-}\mathbb{K}_{123}^{1'2'3'}
(p_{3'}-p_3)^2,
\end{align}
which reduces to Eq.~\eqref{PR} in the main text. For $n=2$
\begin{align}
\dot{E}^R=\frac{\dot{\mathcal{P}}_2}{2m}=-3\frac{\delta\mathcal{T}^R-\delta\mathcal{T}^L}{T^2}
\sum_{++-\atop++-}\mathbb{K}_{123}^{1'2'3'}
(\varepsilon_{p_{3'}}-\varepsilon_{p_3})^2\,.
\end{align}
One additional step is required to obtain Eq.~\eqref{ER}.  Since all
three particles participating in a collision are located near the
Fermi points it means that the characteristic momentum of right
movers is $\sim p_F$ while for the left mover it is $\sim-p_F$. In
contrast, the momenta transferred in a collision $q_i=p_{i'}-p_i$
are much smaller $\sim T/v_F\ll p_F$, which stems from the
temperature smearing of the occupation functions implicit in the
kernel $\mathbb{K}^{1'2'3'}_{123}$. Since $|q_i|\ll p_F$ we
approximate $p_3\approx-p_F$ and linearize the spectrum near the
Fermi points, in particular,
\begin{equation}
\varepsilon_{p_3+q_3}-\varepsilon_{p_3}\approx\frac{1}{2m}[(-p_F+q_3)^2-p^2_F]\approx-v_Fq_3\,,
\end{equation}
which then brings the last expression for $\dot{E}^R$ to the form of
Eq.~\eqref{ER} in the main text.

\begin{figure}
  \includegraphics[width=8cm]{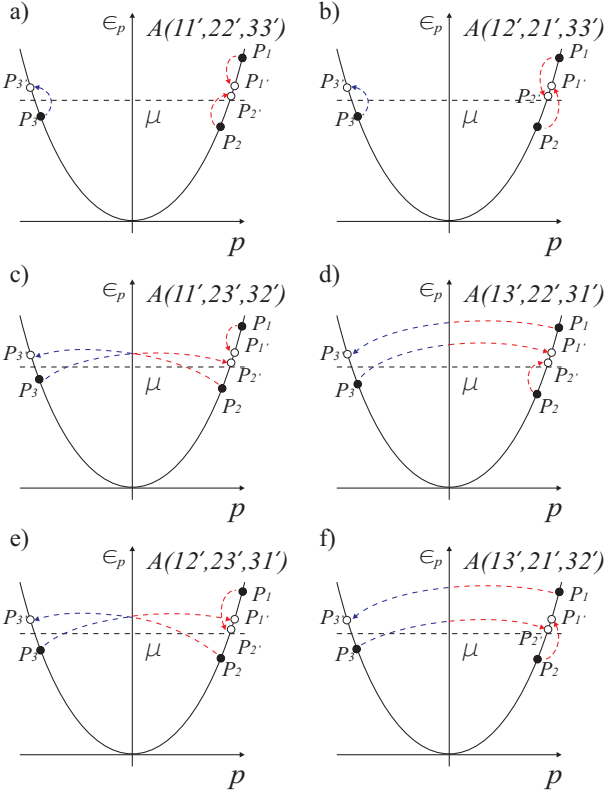}\\
  \caption{(Color online) Direct (a) and five exchange (b)--(f) terms in the three-particle amplitude
  $\mathcal{A}^{1'2'3'}_{123}$ [Eq.~\eqref{Amplitude}] that contribute to the
  finite momentum $\dot{P}^R$ and energy $\dot{E}^R$
  exchange rates between right and left movers.}\label{Fig4}
\end{figure}

\section{Three-particle scattering amplitude}\label{Sec-Appendix-A}

For most of our analysis the detailed form of the scattering rate
entering kinetic equation \eqref{BE} was not important. However, for
the calculation of the microscopic quantities, such as the
scattering lengths $\ell_a$ and $\ell_b$, we need to know the
precise form of the scattering rate $W^{1'2'3'}_{123}$ introduced in
Eq.~\eqref{Coll-Int}. Below we give the details of the structure of
the scattering rate. We start with the golden rule expression
\begin{equation}\label{W}
W^{1'2'3'}_{123}=\frac{2\pi}{\hbar}|\mathcal{A}^{1'2'3'}_{123}|^2
  \delta(E-E')\delta_{P,P'}
\end{equation}
where $\mathcal{A}^{1'2'3'}_{123}$ is the corresponding scattering
amplitude, while the $\delta$-functions impose conservations of the
total energy $E(E')=\sum_{i}\varepsilon_{p_{i}(p_{i'})}$ and total
momentum $P(P')=\sum_{i}p_{i(i')}$ of the colliding electrons. One
should note that in Eq.~\eqref{W} we include $\delta_{P,P'}$ in the
definition of the scattering rate $W^{1'2'3'}_{123}$ rather than the
amplitude $\mathcal{A}^{1'2'3'}_{123}$, which is in contrast to the
usual convention. This step simplifies our notations.

Since the electrons interact with two-particle interaction potential
$V(x)$, the three-particle scattering amplitude is found to the
second order in $V(x)$.  The details of this calculation were
presented in Ref.~\onlinecite{Lunde}. In the case of spinless
electrons the final result reads
\begin{equation}\label{Amplitude}
\mathcal{A}^{1'2'3'}_{123}=\sum_{\pi(1'2'3')}\mathrm{sgn}(1'2'3')A(11',22',33')\,.
\end{equation}
One should notice that Eq.~\eqref{Amplitude} contains the term
$A(11',22',33')$, which is the amplitude of the direct scattering
process [Fig.~\ref{Fig4}(a)], and the terms obtained by the
remaining five permutations of the outgoing momenta, which are the
exchange terms [Figs.~\ref{Fig4}(b)--\ref{Fig4}(f)]. They can be
written compactly for the segment $\Delta x$ of the wire as follows:
\begin{eqnarray}\label{A-aaa}
A(1a,2b,3c)&=&a^{ab}_{12}+a^{ac}_{13}+a^{bc}_{23}\,,\\
a^{ab}_{12}&\equiv&
a^{p_{a}p_{b}}_{p_1p_2}=\frac{1}{(\Delta x)^2}V_{p_{a}-p_1}V_{p_{b}-p_2}\nonumber\\
&\times&
\left[\frac{1}{E\!-\!\varepsilon_{p_1}\!-\!\varepsilon_{p_{b}}
\!-\!\varepsilon_{P-p_1-p_{b}}}\right.\nonumber\\
&+&\left.\frac{1}{E\!-\!\varepsilon_{p_{a}}\!-\!\varepsilon_{p_2}
\!-\!\varepsilon_{P-p_{a}-p_2}}\right],
\end{eqnarray}
where $(a,b,c)$ is a particular permutation of $(1',2',3')$. In
Eq.~\eqref{Amplitude} the notations $\pi(\ldots)$ and
$\mathrm{sgn}(\ldots)$ denote permutations of the final momenta and
parities of a particular permutation. Finally, $V_p$ is the
Fourier-transformed component of the bare two-body interaction
potential. For the calculations we take the Coulomb interaction
between electrons,
\begin{equation}
V(x)=\frac{e^2}{\kappa}\left[\frac{1}{\sqrt{x^2+4w^2}}-\frac{1}{\sqrt{x^2+4d^2}}\right]\,,
\end{equation}
screened by a nearby gate, which we model by a conducting plane at a
distance $d$ from the wire. We also introduced a small width $w$ of
the quantum wire, $w\ll d$, to regularize the diverging short-range
behavior of this potential. This enables us to evaluate the
small-momentum Fourier components $V_p$ of the interaction potential
$V(x)$. To this end, we find in the limit $\hbar/d\ll p\ll \hbar/w$
\begin{equation}\label{V}
V_p= {2e^2\over \kappa}
\ln\left(\frac{p_w}{|p|}\right)\left[1+\frac{p^2}{p^2_w}\right]\,,
\end{equation}
while in the limit of very small momenta $p\ll \hbar/d$
\begin{equation}\label{V-screened}
V_p=\frac{2e^2}{\kappa}\ln\left(\frac{d}{w}\right)
\left[1-\frac{p^2}{p^2_d}\frac{\ln(p_d/|p|)}{\ln(d/w)}\right]\,.
\end{equation}
In the last two equations we introduced the notations $p_w=\hbar/w$
and $p_d=\hbar/d$. We also employed logarithmic accuracy
approximation for $V_p$, meaning that numerical coefficients in the
arguments of the logarithms in Eqs.~\eqref{V} and \eqref{V-screened}
are neglected. In the following discussions we refer to
Eq.~\eqref{V} as the unscreened Coulomb potential and to
Eq.~\eqref{V-screened} as the screened one. The complete expression
for the amplitude \eqref{Amplitude} with the interaction potential
taken in the form \eqref{V} or \eqref{V-screened} is fairly
complicated. However, major simplification is possible by studying
the kinematics of the three-particle collisions, which in a way
allows us to obtain the approximated form of the amplitude for
specific scattering processes, such as the one in
Fig.~\ref{Fig2}(b), which determines the scale $\ell_b$.

It is convenient to label the outgoing momenta as $p_{i'}=p_i+q_i$
for $i=1,2,3$ in order to separate explicitly the momenta $q_i$
transferred in a collision. Momentum conservation then reads
\begin{equation}\label{conservation-momentum}
q_1+q_2+q_3=0\,.
\end{equation}
while energy conservation $E=E'$ can be equivalently rewritten as
\begin{equation}\label{conservation-energy}
2p_1q_1+2p_2q_2+2p_3q_3+q^2_1+q^2_2+q^2_3=0\,.
\end{equation}
At low temperatures, $T\ll\mu$, the Fermi occupation functions
constrain particles participating in the collision to lie in a
momentum strip of the order of $T/v_F\ll p_F$ near the Fermi level.
This means in practice that the typical momentum transferred in a
collision will not exceed $\mathrm{max}\{|q|\}\lesssim T/v_F$. To
leading order in $T/\mu\ll1$ the energy and momentum conservation
requirements for the scattering process in Fig.~\ref{Fig4} can be
resolved by
\begin{subequations}\label{conservation-laws}
\begin{equation}\label{conservation-laws-q12}
q_{1}\approx -q_2+\mathcal{O}\{[(p_1-p_2),q_2]/p_F\}\,,
\end{equation} and
\begin{equation}\label{conservation-laws-q3}
q_{3}\approx\frac{q_1(q_1+p_1-p_2)}{2p_F}+\mathcal{O}\{[(p_1-p_2),q_1]^2/p^2_F\}\,,
\end{equation}
\end{subequations}
where we used $p_1-p_2\sim T/v_F$ and set $p_3\approx-p_F$.  From
this analysis one concludes that energy transfer between the right
and left movers occurs via small portions of momentum $q_3$ exchange
such that
\begin{equation}\label{q123-scale}
\{|q_1|,|q_2|\}\sim T/v_F\,,\quad |q_3|\sim
T^2/v_F\mu\ll\{|q_{1}|,|q_2|\}\,.
\end{equation}
Having two small parameters at hand, $|q_{1}|/p_F\ll1$ and
$|q_3|/|q_1|\ll1$, and accounting for all the exchange
contributions, one can expand the amplitude \eqref{Amplitude} to the
leading nonvanishing order.\cite{ZAK} In the course of this
expansion we observed that exchange contributions result in severe
cancellations between different scattering processes. The result of
the calculations for the model of unscreened interaction \eqref{V}
is
\begin{equation}\label{A-approx-spinless}
|\mathcal{A}^{1'2'3'}_{123}|^2=\left(\frac{2e^2}{\kappa}\right)^4\!\!
\frac{9\lambda_1(k_Fw)}{64\mu^2(\Delta
x)^4}\ln^2\left(\frac{q^2_1}{2p_F|q_3|}\right)
\end{equation}
where the amplitude is written for a segment of the wire of length
$\Delta x$ and the function $\lambda_1(k_Fw)$ was introduced earlier
[see the definition after Eq.~\eqref{l-Q-spinless}]. For the
screened case we find
\begin{eqnarray}
\hskip-.4cm
|\mathcal{A}^{1'2'3'}_{123}|^2\!\!&=&\!\!\left(\frac{2e^2}{\kappa}\right)^4\!\!
\frac{25\lambda_3(k_Fd)}{4\mu^2(\Delta
x)^4}\nonumber\\
\hskip-.4cm
&\times&\!\!\left[\frac{q^2_1}{p^2_F}\ln\left(\frac{p_F}{|q_1|}\right)-
\frac{4q^2_3}{q^2_1}\ln\left(\frac{|q_1|}{|q_3|}\right)\right]^2,
\label{A-approx-spinless-screened}
\end{eqnarray}
where $\lambda_3(z)=z^8\ln^2(1/z)$. Both amplitudes
\eqref{A-approx-spinless} and \eqref{A-approx-spinless-screened} are
written in logarithmic accuracy approximation. As argued above, the
typical scattering processes studied here involve only
small-momentum transfer, of the order of $q\sim T/v_F$. As a result,
from the conditions of applicability of the interaction potential
Eq.~\eqref{V} it follows that the corresponding amplitude
Eq.~\eqref{A-approx-spinless} applies for $T\gg \hbar v_F/d$.
Similarly, the screened interaction potential Eq.~\eqref{V-screened}
and corresponding amplitude Eq.~\eqref{A-approx-spinless-screened}
apply at lower temperatures $T\ll \hbar v_F/d$.

There are several general remarks we need to make regarding the
scattering amplitude in Eq.~\eqref{Amplitude}. It is known from the
context of integrable quantum many-body problems~\cite{Sutherland}
that for some two-body potentials, $N$-body scattering processes
factorize into a sequence of two-body collisions. In the context of
this work, this means that three-particle scattering for the
integrable potentials may result only in permutations within the
group of three momenta of the colliding particles; all other
three-particle scattering amplitudes must be exactly zero for such
potentials. We have checked explicitly that the three-particle
scattering amplitude in Eq.~\eqref{Amplitude} is nullified for
several special potentials: for the contact interaction,
$V_p=\mathrm{const}$, for the Calogero-Suthreland model,
$V_p\propto|p|$, and also for the potential $V_p\propto1-p^2/p^2_0$
which is dual to the bosonic Lieb-Liniger model. Surprisingly, we
have also noticed that the logarithmic interaction potential
$V_p\propto \ln|p|$ gives exactly zero for the three-particle
amplitude in Eq.~\eqref{Amplitude} although we are unaware of any
exactly solvable model for that case. This is the reason to keep the
next leading-order term $\sim(p/p_w)^2\ll1$ in Eq.~\eqref{V}, which
prevents the amplitude in Eq.~\eqref{Amplitude} from vanishing
exactly.

The second set of remarks concern electrons with spin. For the
latter the three-particle amplitude has the same form as
Eq.~\eqref{Amplitude}, but it acquires an additional dependence on
the spin indices:
\begin{equation}\label{Amplitude-Spin}
\mathcal{A}^{1'2'3'}_{123}=\!\!\!\!\sum_{\pi(1'2'3')}\!\!\!\mathrm{sign}(1'2'3')
\Xi^{\sigma_1\sigma_2\sigma_3}_{\sigma_{1'}\sigma_{2'}\sigma_{3'}}
A(11',22',33'),
\end{equation}
where
$\Xi^{\sigma_1\sigma_2\sigma_3}_{\sigma_{1'}\sigma_{2'}\sigma_{3'}}
=\delta_{\sigma_1\sigma_{1'}}\delta_{\sigma_2\sigma_{2'}}\delta_{\sigma_3\sigma_{3'}}$.
One can repeat the expansion of the amplitude for $|q_{1}|/p_F\ll1$
and $|q_3|/|q_1|\ll1$ and observe that due to the spin structure the
exchange terms do not cancel each other. In particular, with the
help of Eq.~\eqref{V} we find the amplitude for the case of the
unscreened Coulomb potential in the form
\begin{equation}\label{A-approx-spinfull}
\!\sum_{\{\sigma\}}\!|\mathcal{A}^{1'2'3'}_{123}|^2\!\!=
\!\left(\frac{2e^2}{\kappa}\right)^{\!4}\!\!\!
\frac{3\lambda_2(k_Fw)}{32\mu^2(\Delta
x)^4}\!\!\left[\frac{4p^2_F}{q^2_1}+\frac{q^2_1}{q^2_3}\right]\!
\ln^2\!\!\left(\!\frac{2p_F}{|q_1|}\!\right),
\end{equation}
which is by a factor of $(p_F/|q_1|)^2\gg1$ larger than
Eq.~\eqref{A-approx-spinless}; $\lambda_2(k_Fw)$ was defined under
Eq.~\eqref{l-Q-spinfull}.

\section{Intra-branch and inter-branch relaxation lengths}
\label{Sec-Appendix-l-Q}

In this appendix we estimate the scattering lengths $\ell_a$ and
$\ell_b$. Our starting point for evaluation of the inter-branch
length $\ell_b$ is the expression
\begin{equation}\label{l-Q-Appendix}
\ell^{-1}_b=\frac{1080\mu^4}{\pi^3T^4}\frac{1}{v_Fk_F\Delta x}
\sum_{++-\atop++-}\frac{(v_Fq_3)^2}{\mu
T}W^{1'2'3'}_{123}\mathbb{F}\{f^0\},
\end{equation}
which follows from Eqs.~\eqref{tau-E} and \eqref{l-Q-def}, where in addition
we introduced the notation
\begin{equation}
\mathbb{F}\{f^0\}=f^0_{p_1}(1-f^0_{p_1+q_1})
f^0_{p_2}(1-f^0_{p_2+q_2})f^0_{p_3}(1-f^0_{p_3+q_3}).
\end{equation}
In view of the kinematic constraints \eqref{conservation-laws-q12}
and \eqref{conservation-laws-q3}, conservation of momentum and
energy in the expression \eqref{W} for the scattering rate
$W^{1'2'3'}_{123}$ can be presented as
\begin{equation}
\delta(E-E')\delta_{P,P'}\approx\frac{1}{2v_F}
\delta\left(q_3-\frac{q_1(q_1+p_1-p_2)}{2p_F}\right)\delta_{q_1,-q_2},
\end{equation}
which eliminates two out of six momentum integrations in
Eq.~\eqref{l-Q-Appendix}. The other four integrals can be completed
analytically with logarithmic accuracy. This amounts to replacing
the weak logarithmic parts of the amplitude in
Eqs.~\eqref{A-approx-spinless}, \eqref{A-approx-spinless-screened},
and \eqref{A-approx-spinfull} by their typical values taken at
characteristic momenta $q_1\sim T/v_F$ and $q_3\sim T^2/v_F\mu$. We
thus treat $\ln(q^2_1/2p_F|q_3|)$ in Eq.~\eqref{A-approx-spinless}
as a constant of order unity and approximate
$\ln(p_F/|q_1|)\simeq\ln(|q_1|/|q_3|)\simeq\ln(\mu/T)$ in
Eqs.~\eqref{A-approx-spinless-screened} and
\eqref{A-approx-spinfull}. After this step we can integrate in
Eq.~\eqref{l-Q-Appendix} explicitly by linearizing the electron
dispersion relation inside the Fermi functions and get
\begin{eqnarray}
&&\hskip-1cm
\sum_{p_1p_2p_3}\!\!q^4_1\mathbb{F}\{f^0\}=\frac{(\Delta
x)^3T}{4h^3v_F}\frac{q^6_1}{\sinh^2\left(\frac{v_Fq_1}{2T}\right)},\\
&&\hskip-1cm \sum_{p_1p_2p_3}\!\!q^3_1(p_1-p_2)\mathbb{F}\{f^0\}=
-\frac{(\Delta
x)^3T}{4h^3v_F}\frac{q^6_1}{\sinh^2\left(\frac{v_Fq_1}{2T}\right)},\\
&&\hskip-1cm \sum_{p_1p_2p_3}\!\!q^2_1(p_1-p_2)^2
\mathbb{F}\{f^0\}=\frac{(\Delta
x)^3T}{6h^3v_F}\frac{q^4_1\left(\frac{7q^2_1}{4}+\frac{\pi^2T^2}{v^2_F}\right)}
{\sinh^2\left(\frac{v_Fq_1}{2T}\right)}.
\end{eqnarray}
For the spinless case and high-temperature regime $T\gg \hbar
v_F/d$, where the Coulomb interaction is unscreened, we obtain
\begin{equation}
\ell^{-1}_{b}\simeq\frac{\lambda_1(k_Fw)}{p^2_FT^4\Delta
x}\left(\frac{e^2}{\hbar\kappa}\right)^4\sum_{q_1}
\frac{q^4_1\left(\frac{q^2_1}{4}+\frac{\pi^2T^2}{v^2_F}\right)}
{\sinh^2\left(\frac{v_Fq_1}{2T}\right)}.
\end{equation}
Note here that we do not keep track of the numerical coefficient in
the expression for $\ell_b$ since within the adopted calculation
with logarithmic accuracy this coefficient is not determined. After
the remaining $q_1$ integration one recovers
Eq.~\eqref{l-Q-spinless}, presented in the main text of the paper.

At lower temperatures $T\ll \hbar v_F/d$, screening effects become
important and one should use Eq.~\eqref{A-approx-spinless-screened}
in the expression for the scattering length
Eq.~\eqref{l-Q-Appendix}. Estimation of $\ell_b$ in this case gives
\begin{equation}
\ell^{-1}_{b}\simeq k_F\lambda_3(k_Fd)(e^2/\hbar
v_F\kappa)^4(T/\mu)^7\ln^2(\mu/T)\,.
\end{equation}

In the spinful case this calculation is completely analogous to the
one above; we just need to use a different expression for the
scattering amplitude. With the help of
Eq.~\eqref{A-approx-spinfull}, which is applicable for the model of
an unscreened Coulomb potential interaction, we get at the
intermediate step with logarithmic accuracy,
\begin{equation}
\ell^{-1}_b\simeq\frac{\lambda_2(k_Fw)}{T^4\Delta
x}\left(\frac{e^2}{\hbar\kappa}\right)^4\!\!\ln^2\left(\frac{\mu}{T}\right)
\sum_{q_1}
\frac{q^2_1\left(\frac{5q^2_1}{2}+\frac{4\pi^2T^2}{v^2_F}\right)}
{\sinh^2\left(\frac{v_Fq_1}{2T}\right)}.
\end{equation}
After the final integration this translates into
Eq.~\eqref{l-Q-spinfull}.

We turn now to discussion of the intra-branch relaxation length
$\ell_a$ introduced in Sec.~\ref{Sec-Preview}.  Unlike the case of
interbranch relaxation, Fig.~\ref{Fig1}(b), here all three colliding
particles are near the same Fermi point; see Fig.~\ref{Fig1}(a). In
this case, the typical momentum change for the three electrons is
the same,
\begin{align}
|q_1|\sim|q_2|\sim|q_3|\sim T/v_F.
\end{align}
At this point we should emphasize that for the processes that
determine the length scale $\ell_b$, a new energy scale $T^2/\mu$
appeared in the problem purely from the kinematic constraints based
on the conservation laws.  This scale determined the typical
momentum transfer of the particle that was alone at one side of the
Fermi surface; see Eqs.~(\ref{conservation-laws-q3}) and
(\ref{q123-scale}).

Another important quantity is the scattering amplitude.  For Coulomb
interaction and for the process where all three particles are near
the same Fermi point, it is a relatively complicated expression, but
similarly to Eqs.~(\ref{A-approx-spinless}) and
\eqref{A-approx-spinless-screened} it depends on momenta only weakly
(logarithmically) for the intra-branch processes.

These two observations help us to estimate $\ell_a$ using the known result
Eq.~\eqref{l-Q-spinless} for $\ell_b$. Namely, by replacing the energy scale
$T^2/\mu$ in $\ell_b$ by $T$, we obtain the estimate
\begin{equation}
\ell^{-1}_{a}\simeq  k_F\lambda_1(k_Fw)(e^2/\hbar
v_F\kappa)^4(T/\mu)^2,
\end{equation}
for the unscreened Coulomb case, $T\gg \hbar v_F/d$.  At lower
temperatures, $T\ll \hbar v_F/d$ it changes to $\ell^{-1}_{a}\propto
T^6$.  It is important to emphasize that regardless of the
interaction model we use there exists a distinct separation between
the scales of the relaxation lengths, namely,
\begin{align}
\ell_a/\ell_b\sim T/\mu\ll1.
\end{align}
This fact justifies our ansatz for the distribution function [see
the discussion after Eq.~\eqref{f-part-eq}].  The detailed
calculation of $\ell_a$ will be presented elsewhere.~\cite{ZAK}


\begin{thebibliography}{99}
\bibitem{Note}
To be accurate we point out that for the noninteracting electrons
$G_0=(2e^2/h)(1+e^{-\mu/T})^{-1}$, where the exponentially small
term comes from the states at the bottom of the band. At low
temperatures $e^{-\mu/T}\ll1$ and thus all such small contributions
are neglected throughout the paper.
\bibitem{Estimate}
For the estimates we use typical parameters from the experiment of
Ref.~\onlinecite{Birge}. In particular, $v_F\sim8\times10^5$m/s,
$k_F\sim10^8$ m$^{-1}$, $\kappa\sim10$, and $w\sim5$ nm. This
translates into $e^2/\hbar v_F\kappa\sim1$ and $\mu=\hbar
v_Fk_F/2\sim25$ meV so that for the temperature $T\sim1$ K the
thermal relaxation length [Eq.~\eqref{l-Q-spinfull}] is of the order
$\ell_b\sim1$ $\mu$m.
\end{thebibliography}
\end{document}